\documentclass[12pt]{emulateapj}
\usepackage[dvips]{color}

\slugcomment{accepted for publication in ApJL}

\shorttitle{Hysteresis of Backflow in  Jets}
\shortauthors{Mizuta et al.}

\begin{document}

\title{Hysteresis of Backflow Imprinted in Collimated Jets}

\author{Akira Mizuta\altaffilmark{1},
Motoki Kino\altaffilmark{2},
and
Hiroki Nagakura\altaffilmark{3}}

\altaffiltext{1}{Center for Frontier Science, Chiba University
Yayoi-cho 1-33, Inage-ku, Chiba, 263-8522, Japan}
\altaffiltext{2}{National Astronomical Observatory of Japan,
Mitaka, 181-8588, Japan}
\altaffiltext{3}{Science and Engineering, Waseda University, 3-4-1 Okubo, Shinjuku, Tokyo 169-8555, Japan}

\begin{abstract}
We report two different types of backflow from jets
by performing 2D special relativistic hydrodynamical simulations.
One is anti-parallel and quasi-straight
to the main jet (quasi-straight backflow), and
the other is bent path of the backflow (bent backflow).
We find that the former appears
when the head advance
speed is comparable to or higher than the local sound speed
at the hotspot 
while the latter appears when 
the head advance speed is slower than the sound speed
at the hotspot.
Bent backflow 
collides with the unshocked jet
and laterally squeezes the jet. 
At the same time, a pair of new oblique shocks are 
formed at the tip of the jet and 
new bent fast backflows
are generated via these oblique shocks.
The hysteresis of backflow collisions 
is thus imprinted in the jet as a node and anti-node structure.
This process also promotes broadening of the jet cross sectional
area and it also causes a decrease in the
head advance velocity.
This hydrodynamic process may be tested by observations 
of compact young jets.
\end{abstract}


\section{INTRODUCTION}
\label{sec:intro}
Recent observations of jets in active galactic nuclei (AGN)
clearly tell us their morphological characters. 
On VLBI scales,
the MOJAVE (Monitoring Of Jets in Active galactic nuclei with VLBA Experiments) 
program clearly shows us a number of actual internal inhomogeneous structures, 
and a growth of the jet's cross-sectional area along its axis
(http://www.physics.purdue.edu/~mlister/MOJAVE/index.html; 
\citet{Lister09}).
Their cross sections are not constant
nor internally uniform.
The importance of jet's internal structures has also been 
used to explain multi frequency spectra
 \citep{Ghisellini05}.
From the standpoint of hydrodynamics,
however,
the generation of internal structures in jets and what determines the
cross sectional area of jets during their evolutions
have not yet been fully investigated.

Most prior work has focused on 
Kelvin-Helmholtz instability (KHI)  
caused by the velocity shear between a jet and  
surrounding matter (e.g., \citet{Hardee88,Perucho06}).
The instability is known as a 
trigger of diamond-shape internal structures caused by oblique shocks 
inside the unshocked jet (e.g, Norman et al. 1984). 
Once the unshocked jet is perturbed by
the KHI in a cocoon which surrounds the jet \citep{Begelman84}, 
then an oblique shock appears inside the unshocked jet and
the velocity component perpendicular to the jet axis
is canceled out through this oblique shock
\citep{Komissarov97,Komissarov98}.

In this Letter, we newly show the importance of backflow.
A backflow is usually
generated at the hotspot of the light jet, where
a light jet is a jet with mass density ($\rho_{\rm j}$)
lower than that of the ambient gas  ($\rho_{\rm a}$),
where the subscripts 'j' and 'a' stand for the jet and
ambient gas, respectively.
Although the existence of backflows is well known 
to previous studies by numerical simulations of the jet propagation
\citep{Norman82,Marti97,Rosen99,
Aloy00,Zhang03},
the importance of the backflow in terms of dynamics 
is not well recognized.
The backflow dynamics are usually ignored since 
the backflow is subsonic flow.
For the first time, it is shown that  backflow is essential
for forming the internal structure and the growth rate of  
the cross-sectional area of jets.

\section{NUMERICAL SET-UP}
\label{sec:method}
We utilize the 
axisymmetric two dimensional special relativistic hydrodynamic
code in cylindrical coordinate ($r\times z$)
developed by one of the authors (AM:
see for details, \citet{Mizuta04,Mizuta06}).
Using the PPM method and TVD Runge-Kutta time integration,
we achieve 3rd order accuracy in terms of
space and time.
Since we do not include any radiative process,
the equations solved in this study are scale free.
The ideal gas equation of state ($p=(\gamma -1)\rho\epsilon$)
is adopted, where 
$p$ is pressure,
$\epsilon$ is specific internal energy, and
$\gamma$ is specific heat ratio which is fixed to $5/3$.
The local sound speed is derived as $c_s=\sqrt{\gamma p/\rho h}=
\sqrt{\epsilon \gamma (\gamma-1) /(1+\epsilon \gamma )}$.

The boundary condition at the cylindrical axis ($r=0$)
 is the reflective condition.
The reflective boundary condition is assumed at $z=0$ except
for the first 10 grid points, i.e., $0<r<1 R_b$ where
a continuous jet is imposed.
The  $R_b$ is the radius of the injected jet from
the computational boundary ($z=0$).
A zero gradient free boundary condition
is assumed at $z=100 R_b$ and $r=r_{\rm max}$.
The unit of the space scale is the beam radius of the injected jet
 ($R_b$).
The unit of the time scale is $R_b/c$,
where $c$ is speed of light.

In this study,
five cases
with different jet velocities ($v_j$) have been 
examined
these are $v_j=0.9c$  (model J09),
$v_j=0.8c$  (J08),
$v_j=0.7c$  (J07),
$v_j=0.6c$  (J06),
and $v_j=0.5c$ (J05).
Corresponding Lorentz factors ($\Gamma$) are 2.3 (J09),
1.7 (J08), 1.4 (J07), 1.3 (J06)
and 1.2 (J05), respectively.
The jet is constantly injected 
from one of the computational boundaries.
Other model parameters are fixed as follows.
The density ratio is fixed to be $\eta\equiv \rho_j/\rho_a=0.1$.
The initial mass density profile of the ambient matter $\rho_a$
is assumed to be uniform.
Mass densities are normalized by the ambient gas, i.e.,
$\rho_a=1$.
Both the jet and the ambient gas are assumed to be cold, i.e.,
$\epsilon_j/c^2 =10^{-2}$ and 
$\epsilon_a/c^2 =10^{-6}$,
where $\epsilon$ is specific internal energy.
Thus specific enthalpy ($h\equiv 1+\epsilon/c^2+p/\rho)$ is 
$\sim 1$.
Corresponding sound speeds of the jet and ambient gas are
$c_{s,j}\sim 0.16c$ and $c_{s,a}\sim 1.6\times 10^{-3}c$, respectively.
The velocity vector of the injected jet is parallel to the cylindrical axis.
We intend to do extensive studies with 
wider parameter space in the near future.

The numerical runs are performed until
the head of the jet reaches the boundaries at $z=100 R_b$.
Since 10 uniform grid points per a $R_b$ are given
in all computational domains
for both radial and $z$ directions,
 there are in total 1000 grid points in the $z$ direction. 
Since the aspect ratio of the distances between
the propagation direction and  lateral direction is different
in each model,
we set different computational domains ($r_{\rm max}$) in $r$ by the models
so that the sideways expanding shock is inside the computational
domain in the last stage of the computation.

\section{NUMERICAL RESULTS}
\label{sec:results}
\begin{figure}
\begin{center}
\rotatebox{90}{\includegraphics[angle=270,scale=0.65]{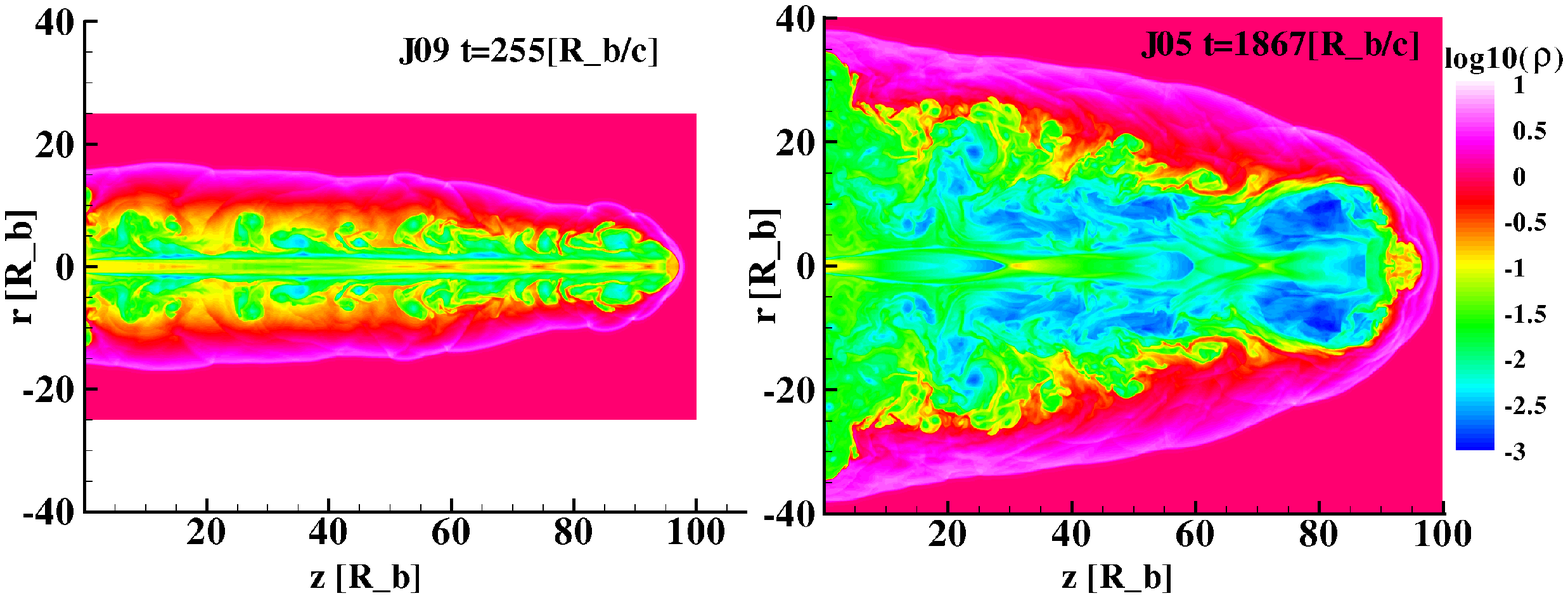}}
\caption{Density contours of the models J05 (left)
and J09 (right) at the end of simulations.
In both cases internal structures can be seen inside the jet.
The jet radius increases at anti-node and decreases at the node
for the case J05,
while the radius of the jet is almost constant along the jet
for the case J09.
\label{density}}
\end{center}
\end{figure}

Here we mainly discuss the results of 
J09 and J05 in order to study differences in backflow dynamics
more clearly. The other three cases will be briefly summarized later on.
Figure \ref{density} shows the  
mass density contours at the final stage
of each model (model J05: $t=1845~R_b/c$, model J09: $t=255~R_b/c$).
Significant differences are found between the two models. They are 
characterized as follows.
\begin{description}
\item
(P1)
Significant nodes/anti-nodes structure 
is generated along the jet in J05. 
The cross sectional radius varies along the jet in J05,
while the cross sectional radius is almost constant along
the jet in J09.

\item
(P2)
The radii of the jet at the terminal shock are about
$1.5~R_b$ in J09,
whilst the radii is  $5~R_b$ 
in J05 which is wider than that in J09.

\item
(P3)
The aspect ratio of the surrounding bow shock 
appearing in J09 is narrower than the one in J05.

\end{description}
In both models,
we observe the backflow which begins the hotspot at the head of the jet.
The temperature in the hotspot is quite high,
since all the jet's kinetic energy is converted to thermal
energy through the terminal shock.
The local sound speeds in the hotspot are 
${c_{\rm s, hs}}\sim 0.5c$ (model J09) and
${c_{\rm s, hs}}\sim 0.3c$ (model J05).
In order to clarify the 
physical reasons for these differences in the density map,
below we examine the time evolution of each model 
in detail.
In advance, we stress that the velocity field maps are the key 
to understanding the above difference because
backflows play an important role in these maps.

Figure \ref{J09velo}(a) shows the contour of the absolute velocity
(color) and velocity vectors (arrows) of
the model J09 in its early phase (at $t=30~R_b/c$).
The jet (red region in contour in Fig.\ref{J09velo}(a))
ends at the terminal shock (at $z=13~R_b$), forming a hotspot.
We find that the jet is surrounded by a layer which
is made of shocked matter escaped from the hotspot.
The velocity of this flow is quite small
and it sometimes shows a negative value measured from the observer 
frame, i.e., $v_z<0$.
This is because  $c_{\rm s,hs}$ and the head propagation speed
$v_h\sim (14R_b)/(30R_b/c) \sim 0.5c$ are very similar,
where  $v_h$ is  the speed of the contact discontinuity
between the shocked jet and the shocked ambient gas
along the jet axis.
The path of this narrow flow
is also almost straight near the head (quasi-straight backflow)
in the late phase of evolution,
(see Fig.~\ref{J09velo}(b)).
Since the straight backflow is unstable for the KHI
between the jet and the backflow,
some vortices are generated along the jet.

Figures \ref{J09velo}(c)-(f) show the evolution of the absolute
velocity contours for model J05.
It is found that the path of the backflow of
model J05 differs dramatically
from that of model J09.
At $t=105R_b/c$, $v_h\sim(13R_b)/(105R_b/c) \sim 0.1c$.
The path of the backflow from the hotspot is not straight but bent
(bent backflow) near the head, (see Fig.~\ref{J09velo}(d)).
The speed of the bent backflow near the hotspot is $\sim 0.25c$.
This bent backflow compresses and piles up the shocked ambient gas.
Then, the backflow makes a half loop path.
Finally, the bent backflow vertically  collides with the main jet
from the side, (see Fig. \ref{J09velo} (d) at $z=7~R_b$).

As shown above, the transition of backflow dynamics 
is caused by different head advance velocity.
The head advance velocity decreases
from model J09 to model J05.
The intermediate models (J08, J07)
show quasi-straight backflow paths, 
while the backflow of the model J06 exhibits
bent path when each head reaches $z\approx 12R_b$,
 see Fig \ref{velocity2}.

\section{PHYSICAL INTERPRETATIONS}
\label{sec:analysis}
In this section, we explain the physical reasons
for the significant difference between the two models.
According to our simulations, the backflow dynamics are
the most important feature in understanding these differences and
we have found mainly two features.
One is the difference in the ratio of 
 $v_{\rm h}$ to ${c_{\rm s, hs}}$.
As we have shown, $v_{\rm h}$ 
is comparable to ${c_{\rm s, hs}}$ in model J09,
whereas the head propagation speed
is slower than local sound speed in model J05.
Since the backflow starts the hotspot,
the speed of the backflow 
at the frame of the hotspot
is close to ${c_{\rm s, hs}}$.
If $v_{\rm h}$ is slower than
the local sound speed, 
the backflow expands sideways.
In the other case, i.e. model J09, the backflow 
goes directly backward and
flows anti-parallel flow to the main jet,
see a schematic figure of the
two types of the backflow path (top panel of Fig.~\ref{backflow}).
By the  comparison of Fig.~\ref{J09velo}(a) and  Fig.~\ref{J09velo}(d),
it is found that
the jets propagated at the same distance from the two models
but along different paths of the backflow,
i.e., quasi-straight backflow and bent backflow.
The bent backflow is generated 
when the condition ${c_{\rm s, hs}}/v_{\rm h}<1-2$ holds.
For ${c_{\rm s, hs}}/v_{\rm h}\sim 2$,
the backflow is clearly separated from the jet.

The other reason is the emergence
of the oblique shock
at the tip of the jet-terminal
which is shown in Fig.~\ref{backflow} (the red line
in the top panel).
The bottom panel of Fig \ref{backflow} shows log-scaled density (top)
and pressure (bottom) contours of the same
model and at the same time of
Fig. \ref{J09velo} (e), i.e., model J05 at $t=270R_b/c$.
The density and pressure discontinuity can be clearly seen
at $15<z/R_b<20, |r|\sim 1R_b$.
This is an oblique shock and is indicated by the dashed line
in the contours (Fig.~\ref{backflow}),
see also Fig. 8(b) \citep{Mizuta04} in which
the oblique shock and fast backflow emerging thorough the shock
can be clearly seen.
When the oblique shock appears,
the backflow comes out not only from the hotspot but also
from the oblique shock, as shown in the shematic figure
in Fig~\ref{backflow}.
A weakly dissipated flow passes through such  an oblique shock
and it becomes bent backflow.
The speed of the bent backflow is $\sim 0.5c$.
Although little attention has been paid to
such a weakly dissipated flow, it is essential
for producing backflows. 
This weakly dissipated flow was seen in some previous studies
i.e., \citet{Saxton02,Mizuta04}.
Since this flow does not pass through hotspots,
the flow directly becomes a backflow, keeping its high velocity.
This flow is separated from the main
jet and draws a half loop in the cocoon.
As a result, a new oblique shock emerges near the terminal,
when the bent backflow vertically hits the unshocked main jet.

P1 and P2 are understood as follows.
Above, we elucidate that the ratio $v_{\rm h}/c_{\rm s,hs}$ 
divides the types of backflows. 
Below, we explain how P1 and P2 are achieved.
When the bent backflow laterally  collides with the unshocked 
jet, the jet is squeezed by the dynamical pinch. 
Thus oblique shocks are generated inside the jet.
Just ahead of the squeezed  node, the jet starts to expand 
like a  nozzle.
This process for generating oblique shocks is completely different
from the previously known one in which internal shocks are caused by
a pressure fluctuation in the cocoon (e.g., Norman et al. 1982).
Since the terminal shock is close to the 
collisional point, 
this oblique shock does not work very well to reconfine the jet.
On the contrary, the cross section of the jet increases at its terminal
as shown in the schematic figure (Fig.~\ref{backflow}).
The wider the cross section becomes, 
the slower the jet advance speed becomes. 
Furthermore, a pair of  oblique shocks represented by the red line 
reaches the terminal of the jet and
connects with the reverse shock.
Once the bent backflow appears from the hotspot,
it is easy to generate 
these oblique shocks,
and new bent backflows intermittently emerge from them 
as discussed above.
The resultant structure therefore  has significant node/anti-node
structures which are the hysteresis of the backflow attacks
as pointed out in P1 and P2.

In J09, $v_{\rm h}$ is almost constant 
throughout the whole calculation.
In J05 $v_{\rm h}$  is also constant
up to $t\sim 200~R_b/c$. After that, the head propagation velocity
begins to decelerate as is shown in the numerical studies of
\citet{Scheck02} and \citet{Mizuta04}.
A large vortex is formed near the hotspot in the simulations. 
The large vortex causes the deceleration of the head propagation.
\citet{Kawakatsu06} analytically obtain  $v_{\rm h}$
taking the growth of head cross sectional area into account
and it reproduces these numerical results well.
Thus, P3 can be simply understood in terms of
previous studies, i.e., the  deceleration of  $v_{\rm h}$.
Finally, it should be noted that
the bent backflow is indirectly linked with
the generation for a large vortex, 
since a weak dissipated flow through an oblique shock
is a key for the appearance of the vortex \citep{Mizuta04}.

So far, we fix $\gamma=5/3$ for simplicity.
 This is inappropriate when the specific internal energy is
 in a relativistic regime ($\epsilon /c^2 \gtrsim 1$),
 and it could be achieved in the hotspot.
 Since the specific heat ratio should be smaller than $5/3$
 in the relativistic regime,
 the actual backflow velocity from the hotspot is slightly slower than
 our estimation.
 Therefore, the transition velocity of backflow dynamics
 is also smaller.
 In order to check the validity of our discussions thus far
 in the real system,
 we examine the case of fixed $\gamma=4/3$
 which corresponds to the relativistic limit.
Three injection velocities, $v_j=0.4c$ (model J04-43), $v_j=0.5c$ (model J05-43),
and $v_j=0.9c$ (model J09-43) are studied.
Figure \ref{velocity3} (a)-(c) show the absolute velocity of these models
in the early phase of the evolution.
 These simulations also show that there are
 two types of backflows,
 the quasi-straight path (models J05-43 and J09-43)
and bent path (models J04-43).
It is verified that the transition velocities are $v_j\sim 0.4c$,
 whereas $v_j\sim 0.6c$ in the fixed $\gamma=5/3$ case.
 That is, the true transition would be a certain value between them.
 Therefore we conclude that
 our simplification does not change the 
 essence of our new findings.

\section{SUMMARY and DISCUSSION}
\label{sec:conclusion}
\begin{figure}
\begin{center}
\rotatebox{90}{\includegraphics[angle=270,scale=0.7]{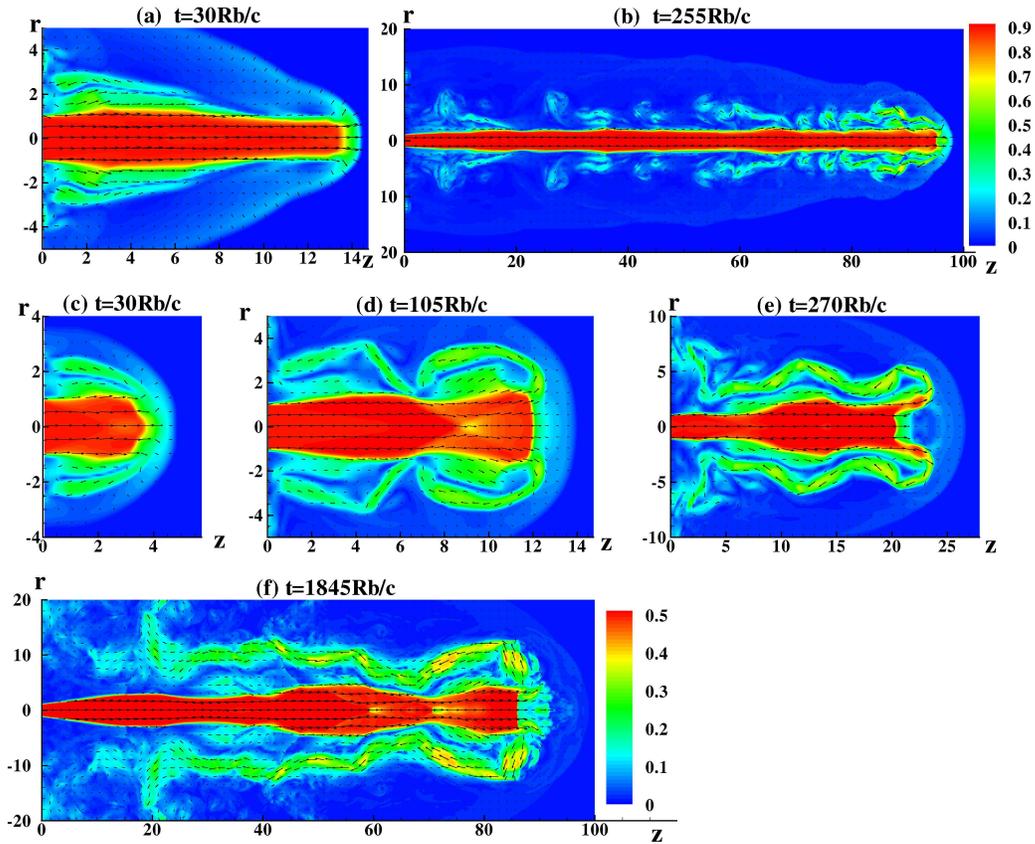}}
\caption{Absolute velocity contours of the model J09 at (a) $t=30$ and
 (b) 255 $R_b$,
and of the model J05 at (c) $t=30$, (d) 105, (e) 270, and (f) 1845 $R_b$. 
The arrows show the velocity vectors.
The color bars are common for (a) and (b), and for (c)-(f).
The backflow is almost straight for the model J09.
Whereas the backflow is bent for the model J05.
A bent backflow attacks the jet from the side (d)-(f).
}
\label{J09velo}
\end{center}
\end{figure}

\begin{figure}
\begin{center}
\rotatebox{90}{\includegraphics[angle=270,scale=0.65]{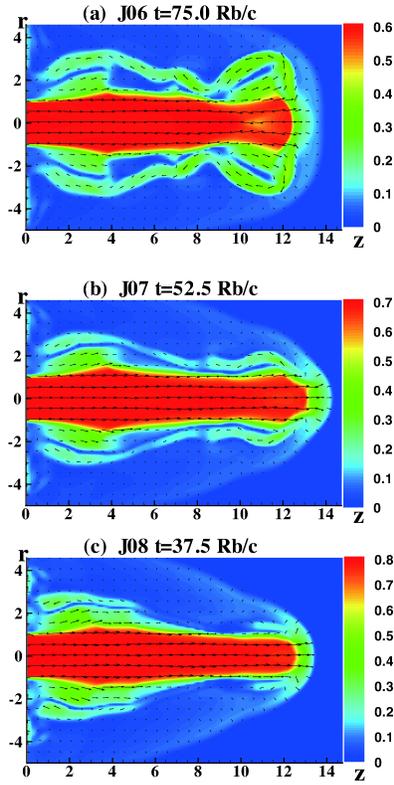}}
\caption{Same as Fig. \ref{J09velo}, but for models
(a)J06 ($t=75~R_b/c$), (b)J07 ($t=52.5~R_b/c$), and (c)J08 ($t=37.5~R_b/c$).
The jet propagates about $12R_b$ in each models,
see also Fig.2(a) and (d) for the comparison.
The path of the backflow is quasi-straight
for models J07 and J08.
On the contrary the path of the backflow is bent for model J06.
\label{velocity2}}
\end{center}
\end{figure}

\begin{figure}
\begin{center}
\rotatebox{90}{\includegraphics[angle=270,scale=1.65]{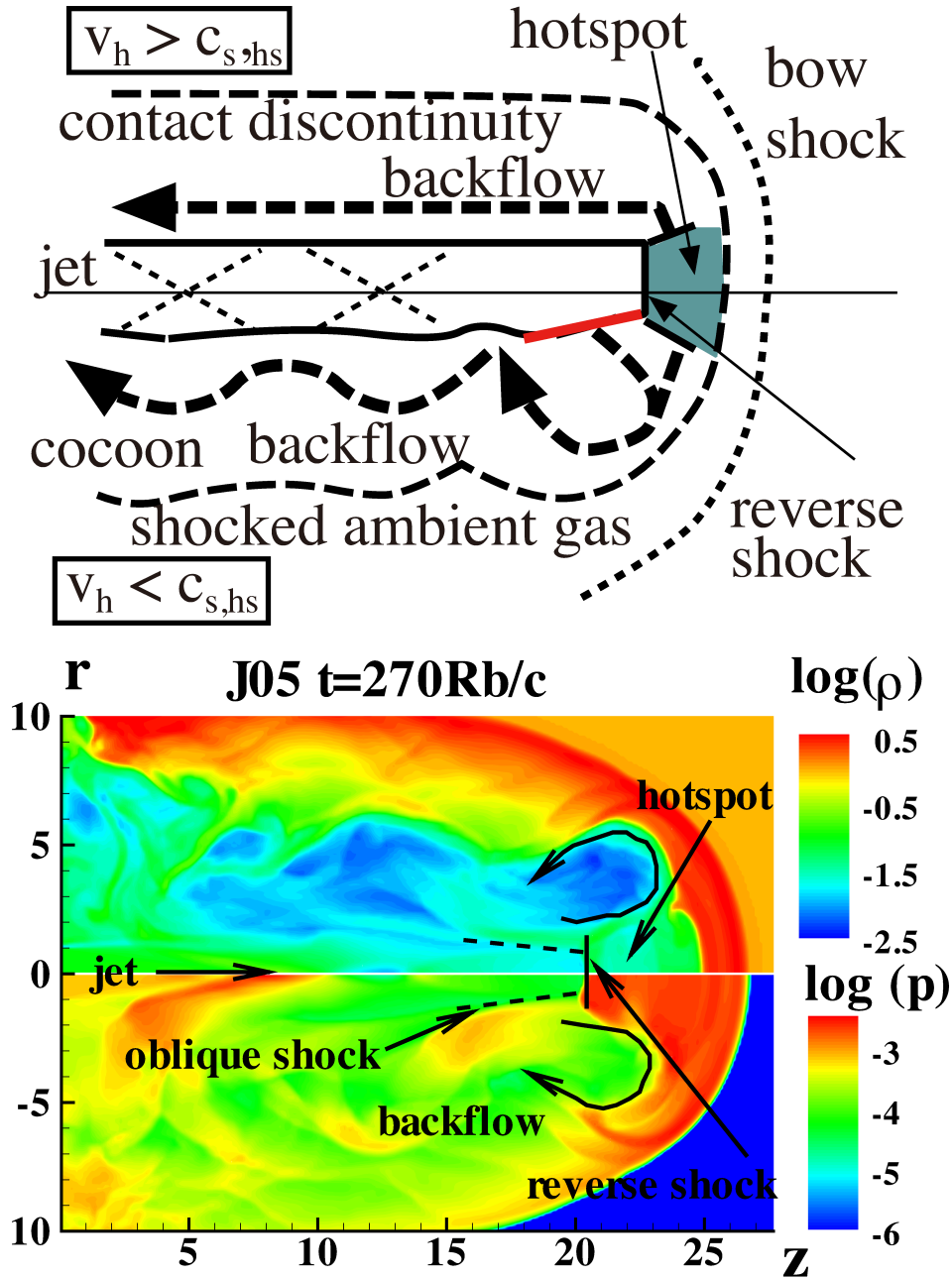}}
\caption{A schematic figure of the path of the jet and backflow (top).
Two cases are shown.
The head speed is faster than the local sound speed at 
the hotspot (upper top) and the head speed is slower than the
local sound speed at the hotspot (lower half).
The path of the backflow is quasi-straight for the former case.
The path of the backflow is bent for the latter case.
The backflow from the hotspot thermally expands sideways,
and becomes bent backflow, interacting with the shocked
ambient gas.
An oblique shock sometimes appears (red line)
at the terminal of the jet which is
another way to trigger bent backflow.
The bottom panel shows log-scaled density and pressure
contours of the same model and at the same time
 of Fig. \ref{J09velo} (e), i.e., model J05 at $t=270R_b$.
Near the terminal shock which is at $z=20R_b$
and perpendicular to the $z$ axis,
a discontinuity indicated by the dashed line can be seen.
This corresponds to an oblique shock shown by the red line in
the top panel.
The backflow mainly comes through this oblique shock at this
snapshot,
see also Fig. \ref{J09velo} (e) in which
a fast backflow starts through this
oblique shock and it becomes a bent backflow.
\label{backflow}}
\end{center}
\end{figure}

\begin{figure}
\begin{center}
\rotatebox{90}{\includegraphics[angle=270,scale=0.65]{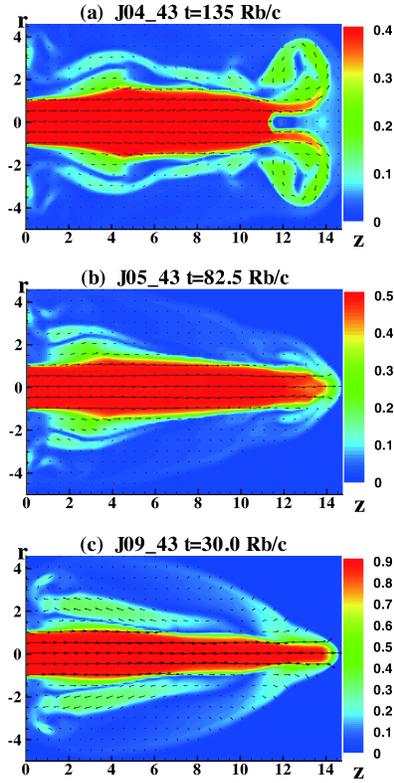}}
\caption{Same as Fig. \ref{J09velo}, but for models
(a) J04-43 ($t=127.5~R_b/c$), (b) J05-43 ($t=82.5~R_b/c$) , and 
(c) J09-43 ($t=30~R_b/c$).
These simulations are done with fixed specific ratio $\gamma=4/3$.
The path of the backflow is quasi-straight one (model J05-43 and J09-43).
On the contrary the path of the backflow is bent one for model J04-43.
The transition of backflow property occurs at slower injection 
velocity than that for the models done with fixed specific heat ratio 
$\gamma=5/3$.
\label{velocity3}}
\end{center}
\end{figure}

In this work, we show two types of backflow dynamics,
quasi-straight path and bent path.
The difference can be clearly seen in the early phase of the
dynamics, (see Figs. \ref{J09velo} (a) and (c)).
When the head propagation velocity is smaller than
local sound speed at the hotspot,
a bent backflow appears.
The bent backflow goes backward and sometimes attacks
the jet from the side.
When the jet is dynamically attacked by the bent backflow,
the cross section of the jet increases, making an oblique shock.
At the terminal of the jet,
an oblique shock and the terminal reverse shock
connects, making a weak dissipated flow into the cocoon
through the oblique shock.
Such backflow also becomes a bent one
and appears repeatedly.
The transition of two types of backflow dynamics
occurs by decreasing the head propagation velocity.

Compact Symmetric Object
(CSO) would be one of the best targets for testing our findings,
since CSOs are newly born AGN jets with radio lobes.
For instance, one of the smallest CSO radio lobes 
3C 84 \citet{Asada06} with the 
size of $\sim $ 10 pc may be 
a good target to explore the backflow physics investigated in this work. 
Observations with unprecedented high spatial resolution 
which will be 
provided by VLBI Space Observatory Programme 2 (VSOP-2)
in the near future appear to provide a promising
means of exploring these backflows.

We add brief comments on the relevance to 
gamma-ray bursts (GRBs). 
Hot jets are expected in GRBs and
it is intriguing to see how backflows act when jets are hot.
Since the GRB jet from collapsars is formed deep inside of the progenitor,
the jet should drill the high density stellar envelopes to
reach the progenitor surface (e.g., \cite{Lazzati09}).
For example, the bent path of the backflow from the head of the GRB jet
is indeed seen in Figures 2, 3 and 4 in \citet{Zhang04},
during its propagation in the high density progenitor envelope.

\acknowledgments

We are grateful to an anonymous referee for beneficial comments.
This work is partly supported by the Grants-in-Aid of the Ministry
of Education, Science, Culture, and Sport (20041002, 19540236, 21018002 A.M.).
This work was carried out on 
NEC SX8, at Yukawa Institute for Theoretical Physics, Kyoto University,
on the Space Science Simulator (NEC SX6), at the Japan
Aerospace Exploration Agency, and on XT4 system and NEX SX9, at the Center for
Computational Astrophysics, National Astronomical Observatory of
Japan.

\clearpage

\clearpage

\clearpage

\clearpage

\end{document}